# Stabilization of phase-pure rhombohedral HfZrO4 in Pulsed Laser Deposited thin films

*Laura Bégon-Lours\*, Martijn Mulder, Pavan Nukala, Sytze de Graaf, Yorick Birkhölzer, Bart Kooi, Beatriz Noheda, Gertjan Koster and Guus Rijnders*

Dr. Laura Bégon-Lours*,
Inorganic Materials Science, University of Twente, 7500 AE, The Netherlands
E-mail: l.c.y.begon-lours@utwente.nl
Martijn Mulder,
Inorganic Materials Science, University of Twente, 7500 AE, The Netherlands
Dr. Pavan Nukala,
Zernike Institute of Advanced Materials, University of Groningen, 9747 AG, The Netherlands
Sytze de Graaf,
Zernike Institute of Advanced Materials, University of Groningen, 9747 AG, The Netherlands
Yorick Birkhölzer,
Inorganic Materials Science, University of Twente, 7500 AE, The Netherlands
Prof. Bart Kooi
Zernike Institute of Advanced Materials, University of Groningen, 9747 AG, The Netherlands
Prof. Beatriz Noheda
Zernike Institute of Advanced Materials, University of Groningen, 9747 AG, The Netherlands
Prof. Gertjan Koster
Inorganic Materials Science, University of Twente, 7500 AE, The Netherlands
Prof. Guus Rijnders
Inorganic Materials Science, University of Twente, 7500 AE, The Netherlands

**Abstract**: Controlling the crystalline structure of Hafnium Zirconate and its epitaxial relationship to a semiconducting electrode has a high technological interest, as ferroelectric materials are key ingredients for emerging electronic devices. Using Pulsed Laser Deposition, a phase pure, ultra-thin film of $Hf_{0.5}Zr_{0.5}O_2$ is grown epitaxially on a GaN (0001) / Si (111) template. Since standard microscopy techniques do not allow to determine with certitude the crystalline structure of the film due to the weak scattering of oxygen, differentiated differential phase contrast (DPC) Scanning Transmission Electron Microscopy is used to allow the direct imaging of oxygen columns in the film. Combined with X-Rays diffraction analysis, the polar nature and rhombohedral R3 symmetry of the film are demonstrated.

# 1. Introduction

The integration of ferroelectric materials on semiconductors paves the way to the development of novel technologies based on the combination of functionalities of two different materials systems. First, GaN-based high electron mobility transistors (HEMTs) for high power electronics show excellent performances thanks to a wide band-gap and high electron drift velocity. The use of a ferroelectric gate further increases the performance (i.e. by increasing the threshold voltage or the ON/OFF ratio) of such transistors.[1–3] However the integration of such ferroelectric materials (usually perovskites[2,4–7]) on GaN is challenging due to the large lattice mismatch between the two structures and a buffer layer, such as MgO,[2,7] $Al_2O_3$,[6] or $TiO_2$,[8] is commonly used to allow for the mismatch. Direct integration of a fully epitaxial ferroelectric material on GaN is the promise of further improvement for such devices. Furthermore, the remanence and electric field control of the polarization makes ferroelectric materials ideal candidates for non-volatile memories and in-memory computing.[9] Specifically, $HfO_2$-based ferroelectric structures on silicon[10] or transparent conducting oxides[11] show excellent memristive behaviors and low-voltage operation thanks to its high-dielectric constant. Layers of doped-Hafnium have been extensively studied since 2011, when a ferroelectric phase of this common dielectric material was first discovered[12] in films synthesized by Atomic layer Deposition (ALD). These films exhibited ferroelectricity only in the nanoscale and metastable polar polymorphs were established to be responsible for this behavior. The lowest-energy orthorhombic (o-) phase was widely observed in many studies. The exploration of the conditions and mechanisms to obtain the o-phase is on-going, in particular the role and the nature of the top and bottom electrodes: epitaxial strain,[13] mechanical strain,[14] oxidation of the electrode,[15,16] the composition of the film[15,17–19] and the thermal annealing conditions.[20–22] ALD however always results in polycrystalline films, with an overall polarization smaller compared to the polarization of single grains, in addition to the need of a wake-up process. Other techniques such as sputtering,[23] MOCVD[24] (including on GaN[1]) or CSD[25] have been explored, similarly resulting in polycrystalline films. On the contrary, Pulsed Laser Deposition (PLD) is a vapor phase technique that allows the direct epitaxial growth of thin films. It has been recently used for the fabrication of HZO thin films on perovskite electrodes[26–29] and structural analysis of films grown on LSMO/STO (001) substrates demonstrated the occurrence of a novel phase with rhombohedral (r-) symmetry.[26]

The r- and o- phases had in common that they only occurred for film thickness below 10 to 15 nm.[21,26] As the thickness of the film increased, the fraction of grains crystallizing in the bulk monoclinic phase increased, limiting the applications requiring ferroelectricity. Despite ab-initio studies of HfO$_2$ with several dopants,[30] it is still unclear what conditions led to the stabilization of which polar phase, pointing to the need for the systematic study of high-quality films under various conditions (substrates, orientation, etc.). As the thickness of the film decreased (below ~5 nm), the ferroelectricity vanished,[31,32] either because of depolarizing field at the interface with the electrodes,[33,34] or because of an interfacial layer in a different phase.[26] Controlling the formation of ferroelectric phases of hafnium zirconate is necessary for the fabrication of novel non-volatile memories, as it is a candidate for the fabrication of Negative Capacitance Transistors,[35] Ferroelectric Field-Effect Transistors,[36–39] or Ferroelectric Tunnel Junctions[31,40] devices.

In this context, here we explore the epitaxial growth of HZO by Pulsed Laser Deposition on a template with a trigonal symmetry, Gallium Nitride buffered Si (GaN (0001)). In this research, using RHEED, XRD and aberration corrected Scanning Transmission Electron Microscopy (STEM), we will show that the HZO film crystallized in a rhombohedral symmetry, is phase-pure and is [111] oriented. Fundamentally, this is the first demonstration of a pure r-phase on a non-perovskite substrate and provides more clues to its preferential stabilization. From an application stand-point, our work demonstrates that oriented polar r-phase can be seamlessly integrated onto Si-based substrates.

2. Results

2.1. Epitaxy of the HfZrO$_4$ film on GaN

We first analyze the growth mode during the deposition. First, the main diffraction peaks in the RHEED pattern of the GaN substrate and the azimuthal angles are measured, as shown in **Figure 1** a) and b). The substrate is then heated, and the intensity of the main diffraction peak is monitored during the growth. It shows a drop in intensity in the very early stage of the deposition, followed by a recovery as the thickness of the film increases. In addition, the morphology of the film reproduces the features (terraces) of the GaN template (an AFM image can be found in **Figure S1**), which is indicative of a steady state (or step-flow-like) growth mode for the HZO film. After the deposition and cooling down, photographs of HZO diffraction patterns are taken. We observe that films of 3 to 11 nm present very similar patterns with comparable d-spacings, indicating homogeneous growth. The homogeneity of the film after cooling down is confirmed by EDX mapping of the chemical elements in the film (see Figure S2). The RHEED patterns of the film show a single crystalline phase, with a six-fold symmetry. **Figure 1** compares RHEED patterns of the HZO film (c,d) and of the GaN substrate (a,b): the main symmetry axis is found at the same azimuthal angle the main diffraction axis $[10\bar{1}0]$ and $[11\bar{2}0]$ of the GaN template, demonstrating an epitaxial relationship between the film and the substrate.

## 2.2. Determination of a polar, rhombohedral phase of HZO

Standard XRD analysis was carried out on the HZO film. The thickness of the HZO film was determined to be 58,3 Å ≈ 20 × 2.96 Å by X-Ray Reflectometry. Symmetric scans were carried out around the Si (222) peak of the substrate, shown in **Figure 2** a). The GaN epitaxial film signature consists of three peaks for the (0001) reflection. The peaks at 2θ = 30.20° (plain arrow) and 2θ = 32.30° (dashed arrow) originate from the HZO film. The out-of-plane lattice parameter associated with the peak at 2θ = 30.20° is 2.96 Å. Thus the position of the feature at 2θ = 32.30° matches the expected position for a Laue fringe[41] of 20 diffracting planes separated by 2.96 Å, confirming that the film is highly oriented and structurally perfect. The inset of **Figure 2** a) shows the rocking curve around HZO peak, showing a full width at half maximum of 0,26°. Note that in the literature, a strong peak at 2θ = 30.20° is generally attributed to either the orthorhombic or the tetragonal phase of HZO but further analysis below will show it is also compatible with the rhombohedral phase.

**Figure 2** b) shows pole figure of the HZO film taken at a tilt of χ ≈ 71°. It reveals six peaks at separated in φ by 60°. This is consistent with the existence of two domains (D1 comprising of P1-P3 and D2 comprising of P4-P6) rotated by 180° with respect to the [111] growth axis. In comparison, pole figure of the GaN/Si substrate only show three poles, consistent with the trigonal symmetry of GaN(0001) and Si(111). **Figure 2** c) shows symmetric scans taken at these poles. Scans across P1, P2 and P3 (belonging to one domain), show a peak at 2θ = 30.45° whereas the out-of-plane diffraction peak was at 2θ = 30.20°. Such a 3:1 multiplicity is a strong signature of a rhombohedral space group.

We then analyzed the local structure of our films using STEM imaging. HAADF-STEM images were taken from the cross-section of the film, revealing a homogeneous film with two types of domains (labelled R1 and R2 **on Figure 3** a)) Local Fast Fourier transforms of the image (**Figure 3** b,c) show that the domains are (111) oriented, and are rotated 180 deg with respect to the [111] direction, consistently with our pole figure analysis. The corresponding plane spacing measured by XRD are $d_{111}$ = 2.96 Å and $d_{1\text{-}11}$ = 2.93 Å (error: ±0.02 Å). Closer analysis of the images indicate that the R grains relax close to the surface. The *d*-spacings from STEM analysis are listed in **Table 1**.

HAADF image shown in **Figure 3** d) from a single domain shows alternating contrast in the cationic columns along the <112> in-plane. This is consistent with our multislice HAADF simulations of *R3* and *R3m* phases (and not the o-phase) obtained at sample thickness of 20 nm (see inset). Thus, comparing HAADF simulations of known HZO phases to the experimental data confirms the rhombohedral symmetry of the lattice.

To further determine the precise symmetry of the r-phase, we performed oxygen column imaging via differential phase contrast (DPC) STEM, as shown in **Figure 4** a) and b). The differentiated DPC (dDPC) images were simulated for *R3* and *R3m* phases (structures obtained from Wei *et al.*[26]) through multislice simulations (**Figure 4** c). In the *R3m* phase, O-Hf-O//O-Hf-O columns (along the direction indicated in **Figure 4** c, top, by the red line) are collinear, whereas in the *R3* phase, the loss of mirror symmetry results in the loss of this collinearity when observed along the [110] zone (**Figure 4** c, bottom, red and yellow lines). In our experimental dDPC (**Figure 4** b) also there is also no collinearity, suggesting that *R3* is a better match for our phase than *R3m*. The simulated unit cell for the *R3* phases, shown in **Figure 4** d, also shows better resemblance to the dDPC image than the *R3m* phase.

Next, we estimated the polarization of the R3 phase from our dDPC images. The position of the atomic column in one unit cell ($HfO_2$) was assigned as the position corresponding to the maximum intensity of the column. These columns are overlaid on the top of the image (**Figure 4** b). The center of mass of the cationic columns ($V_c$) and anionic columns ($V_a$) was computed across four different unit cells, and displacement ($d = V_c-V_a$) of 8.5-9 pm was measured in the [111] direction (consistent with the rhombohedral symmetry). The polarization is then roughly estimated using the equation $P = qe * d/V$, where $q = 5$ represents the Born effective charge on the cation in HZO,[42,43] $e = 1.602*10^{-12}$ C, $V$ is the unit-cell volume (~395 Å$^3$), to be 1.6-1.9 µC.cm$^{-2}$, pointing out-of-the film (along [111]).

## 2.3. Determination of the unit cell parameters

Further XRD analysis were carried out to confirm the symmetry of the bulk film and determine the lattice parameters of the cell. Considering a R3 lattice, (111) oriented, IP-GIXRD analysis with a large area 2D detector, of 14.5° in 2θ and 7° in γ, was carried out to access the (-220) and (-440) diffraction peaks. The (-220) peak (resp. (-440)) had a six-fold symmetry in φ and was centered around 2θ = 50.8° (resp. 113.0°), which corresponds to a *d*-spacing of $d_{-220}$ = 1.80 Å (φ-scans can be found in **Figure S3**). This value is in accordance with the *d*-spacing measured with quantitative analysis of the RHEED images for the surface (1.81-1.83 Å) found for the high-symmetry direction parallel to the [11-20] axis of GaN.

Finally, off-axis measurements were carried on to measure the d-spacing along the (200), the (311) and the (042) directions. Diffraction peaks with a six-fold symmetry in φ were obtained for χ = 55.4° and 2θ = 35.4°, indicating $d_{200}$ = 2.53 Å. Similarly, diffraction peaks with a six-fold symmetry in φ were obtained for χ = 30° and 2θ = 60.2°, indicating $d_{311}$ = 1.54 Å. Additional diffraction peaks with a six-fold symmetry in φ were obtained for χ = 39.2° and 2θ = 85.3°, indicating $d_{042}$ = 1.14 Å.

Assuming $d_{(0-11)}^2 = d_{(001)}^2 + d_{(010)}^2 - 2 d_{(001)} \cdot d_{(010)} \cos(180 - \alpha)$, with α the rhombohedral angle, which is a valid approximation for α<1°, we estimate from the values measured by STEM (see table 1): α = 89.5 ±0.2°.

These values differ by a few percents of the values predicted by DFT for the R3 phase, with a rhombohedral angle α = 88.6°. By changing α from 88.6° to 89.4° and assuming $a \sim d_{(001)}$ = 5.07 Å, the computed values of the d-spacing of the planes explored by XRD and RHEED matches the experimental values, as well as the angles between the measured planes and the (111) direction. The projection of the calculated unit cell using the VESTA[44] software are shown in **Figure S4**.

Furthermore, the r-phase HZO reported by Wei et al.,[26] on perovskite electrodes had a larger rhombohedral distortion with α = 88.6°. α in rhombohedral ferroelectric materials is directly correlated to the values of Pr.[45] Thus relatively lower values of Pr in this work is a result of lower rhombohedral distortion in contrast to the r-phase grown on perovskite electrodes (STO//LSMO).

3. Conclusion

By a rigorous structural analysis combining STEM, XRD and RHEED, we demonstrated the epitaxial growth by PLD of a phase-pure rhombohedral phase of HZO on GaN (0001) belonging to the R3 space group and determined its lattice parameters. The found R3 space group (point group 3), enantiomorphic and polar, is compatible with ferroelectricity. This is the first demonstration of the epitaxial growth of a polar phase of Hafnium Zirconate on the large-band gap material Gallium Nitride.

## 4. Experimental Section

An $Hf_{0.5}Zr_{0.5}O_2$ (HZO) target was synthesized by pressing $HfO_2$ and $ZrO_2$ powders, followed by an annealing of 4 hours at 1400°C in air. The surface of GaN/Si (111) substrates was decontaminated by a dip of 30 s in HF1% followed by a 3' etching in HCl (36%), prior to introduction in the PLD chamber. HZO thin films of various thicknesses were grown by Pulsed Laser Deposition method, operating a KrF excimer laser ($\lambda$ = 248 nm) at 2 Hz, with a fluence of 1,6 J.cm$^{-2}$. The target-substrate distance was 50 mm, the substrate was heated to 750°C and the background gas was oxygen at a pressure of 0.1 mbars. Prior to the deposition of the HZO film, the main diffraction axis $[10\bar{1}0]$ and $[11\bar{2}0]$ of the GaN substrate were subsequently aligned with respect to a RHEED beam operating at 30 kV and photographs were taken. During the deposition, the RHEED beam was aligned with the $[10\bar{1}0]$ axis of the GaN substrate. After deposition, the films were cooled down to room temperature in the same background gas, at a ramp rate of 20°C.min$^{-1}$. In situ XPS analysis was carried out in a base pressure of 5E-11 mbars, using a monochromatic Al K$\alpha$ source with a kinetic energy of 1486.7 eV, and a 7-channel analyzer.

For the symmetric XRD scans the Panalytical XPert3 Pro MRD system was used with Cu-k$\alpha$ radiation ($\lambda$ = 1,5406 Å) with 45 kV and 40 mA in line-focus mode. The setup included a PIXcel 3D detector on the diffracted beam side and a 4-crystals Ge(220) monochromator on the incident beam side. For the IP-GIXRD measurements, a Bruker D8 Discover diffractometer was used with a rotating anode microfocus source. On the incident beam side Montel mirror optics with focused beam in the vertical direction, and a parallel beam in the horizontal direction, were used. No monochromator was used for these measurements. A 200 µm double pinhole collimator, which reduces beam divergence to better than 6 mrad to produce a quasi-point source was placed. On the diffracted beam side, no secondary optics were used. The detector used was an Eiger 2R 500K large area detector, with a range of 14.5° 2θ and 7° in γ at a sample detector distance of 290 mm. Pole figures and subsequent pole-slicing were obtained in a point focus mode.

For off-axis XRD measurements both the Bruker D8 diffractometer and the Panalytical XPert3 Pro MRD system were employed. In the MRD, the PIXcel 3D detector was used, with on the incident beam side a 1/2° divergence slit and a 4 mm mask and no monochromator. In this system, the incident angle was changed to align the crystal planes for the diffraction measurements. In the Bruker D8, the angle χ of the sample was tilted to align the planes to measure them.

Electron transparent lamellae for STEM measurements were made through the standard focus ion-beam procedure (Thermofischer Helios G4 CX). The STEM measurements were conducted on a Themis-Z from Thermofischer Inc in STEM imaging mode. HAADF-STEM images were obtained with the detector with collection angle range 65-200 mrad. dDPC images were obtained using a detector segmented in four quadrants that each span the collection angle range of 8-30 mrad, and the contrast has been inverted for better visibility. The High tension was 300 kV and the beam convergence angle was 23.1 mrad. The measured screen current was 52 pA. Multislice image simulations of $HfO_2$ were performed using Dr. Probe software[46]. For the simulations the calibrated experimental values of the beam semi-convergence angle and detectors collection angles were used, and all aberrations, except for defocus, were set to zero. The $HfO_2$ crystal was divided in four equally thick slices that each contain one atomic plane. The dDPC images where computed based on the work of Lazić et al.[47] The final simulated images were convolved with a two-dimensional Gaussian with a full-width at half maximum of 70 pm, to account for the finite probe size.

**Acknowledgements**

L.B.-L. acknowledges the funding received from Netherlands Organisation for Scientific Research NWO under grant agreement no. 718.016.002 ("TOP-PUNT") P.N. acknowledges the funding received from the European Union's Horizon 2020 research and innovation programme under Marie Sklodowska-Curie grant agreement no. 794954 (nickname: FERHAZ).

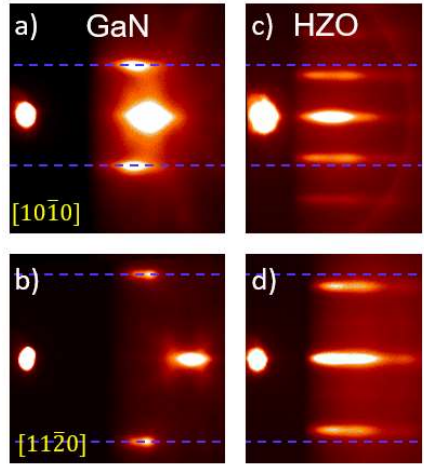

**Figure 1.** RHEED diffraction patterns of GaN (a,b) and HZO (c,d) along the [10-10] direction of GaN (a and c) and along the [11-20] (b and d), showing the epitaxy of HZO on GaN. The blue lines are guides to the eye.

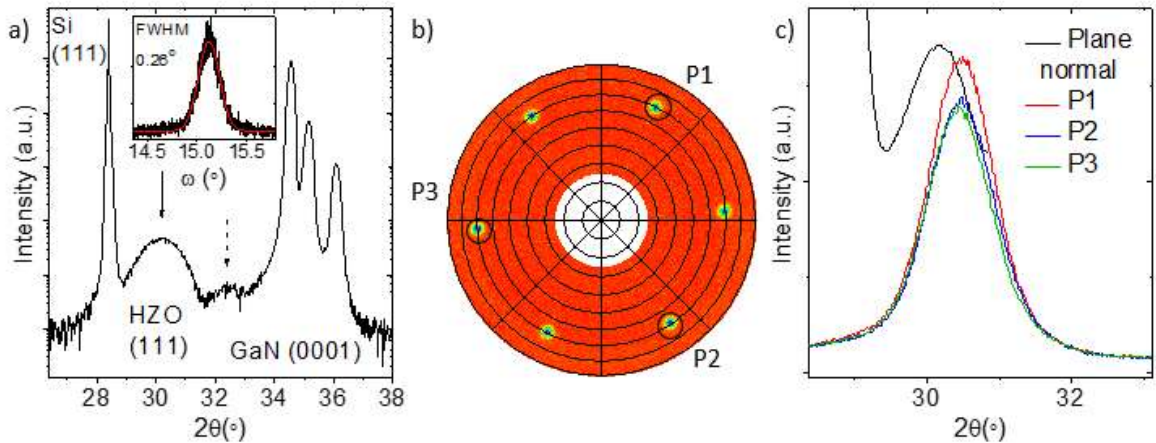

**Figure 2.** a) Symmetric XRD scan of a Hf$_{0.5}$Zr$_{0.5}$O$_2$ film with a thickness of 5.9 nm. The plain arrow indicates the HZO (111) diffraction peak and the dashed arrow a satellite Laue fringe. The inset shows a rocking curve around HZO diffraction peak. b) Pole figures of the same film with six peaks at $\chi \approx 71°$ labeled with P1-P6. c) Black: symmetric XRD scans of the 111 peak, Red/Blue/Green: GIXRD scans of peaks P2, P4, P6.

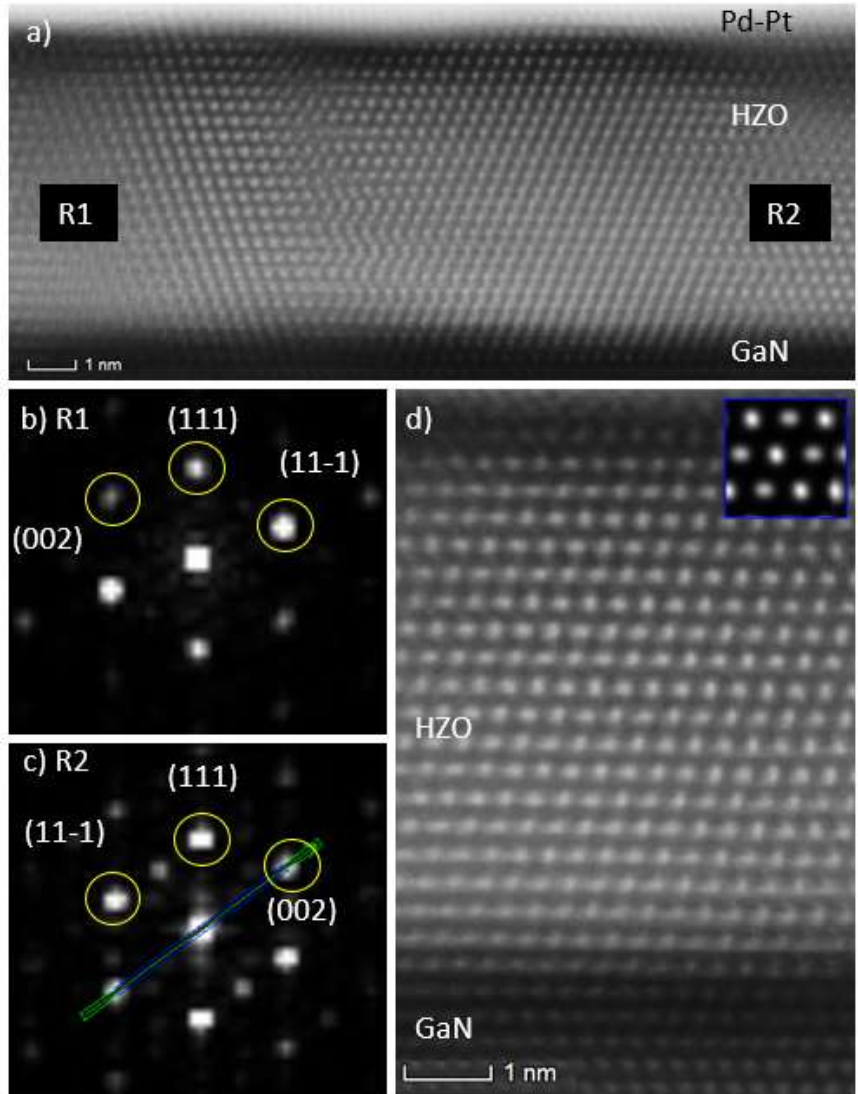

**Figure 3.** a) cross-sectional HAADF-STEM image of a 5.9 nm Hf0.5Zr0.5O2 sample. The different layers are labeled with their respective material. In the HZO layer, the R1 and R2 domains have been indicated. b) and c) show the Fourier transforms of the R1 and R2 domains, in which the diffraction spots are labeled. d) HAADF-STEM image of a R2 domain. The inset shows the simulated HAADF image for the r-phase.

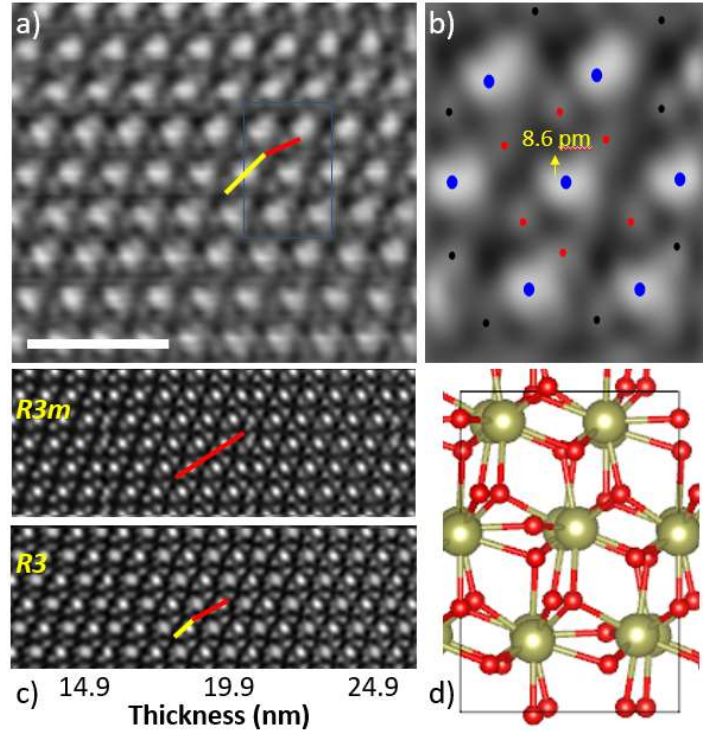

**Figure 4.** a) dDPC image of a HZO film, scale 1 nm. The blue box is enlarged in b). The blue (resp. red) spots represent the position of the Hf/Zr (resp. Oxygen) atoms, with an overall displacement on 8.6 pm. c) Simulated dDPC images for the R3m and R3 phases. The yellow and red lines emphasize the best match between experimental data and R3 phase. d) simulated R3 unit cell.

**Table 1.** lattice plane spacing measured by STEM and XRD.

| Å | R1 | R2 | R1-relaxed | XRD |
|---|---|---|---|---|
| d(111) | 2,95-3,02 | 2,96-2,98 | 2,92-2,95 | 2.96 |
| d(11-1) | 2,91-2,93 | 2,87-2,92 | 2,93-2,95 | 2.93 |
| d(001) | 5,06-5,07 | 5,05-5,07 | 5,05-5,11 | 5.06 |

**Supporting Information**

**Stabilization of phase-pure rhombohedral HfZrO4 in Pulsed Laser Deposited thin films**

*Laura Bégon-Lours\*, Martijn Mulder, Pavan Nukala, Sytze de Graaf, Yorick Birkhölzer, Bart Kooi, Beatriz Noheda, Gertjan Koster and Guus Rijnders*

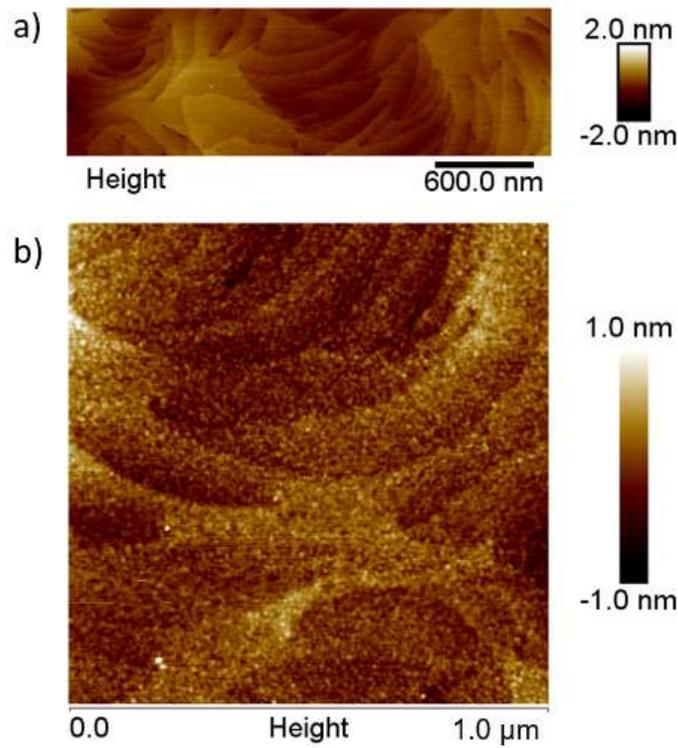

**Figure S1:** AFM topography images of a) a bare GaN substrate showing typical terraces and b) a 6 nm HZO film.

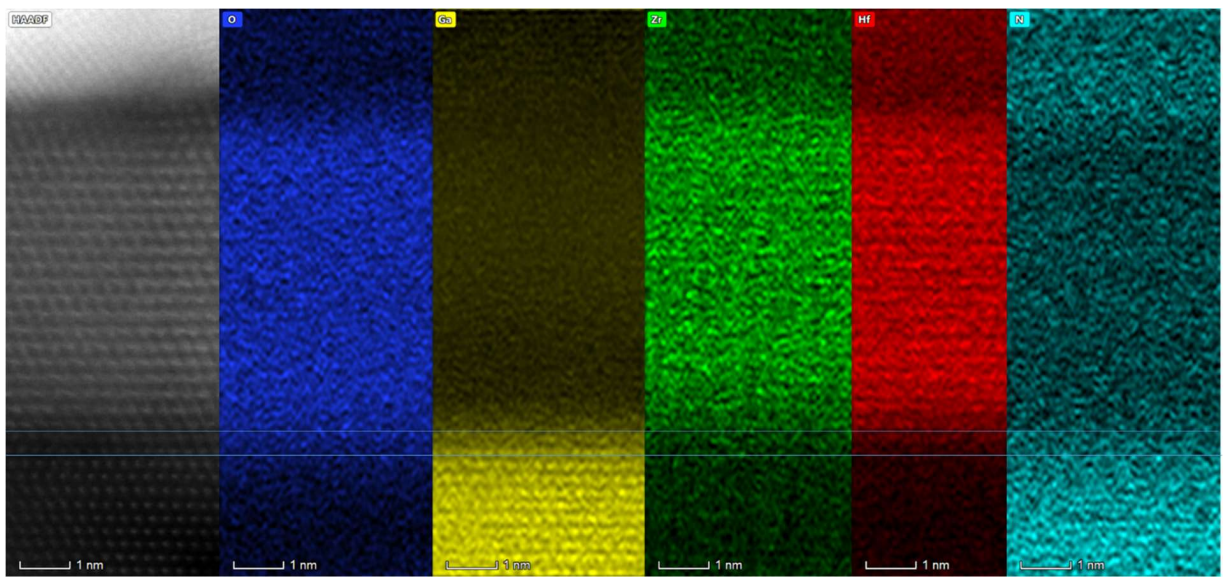

**Figure S2:** HAADF-STEM image (left) and corresponding EELS-maps of the HZO thin films for the elements oxygen (blue), gallium (yellow), zirconium (green), hafnium (red) and nitrogen (teal). The horizontal blue lines indicate the interfaces between the different layers.

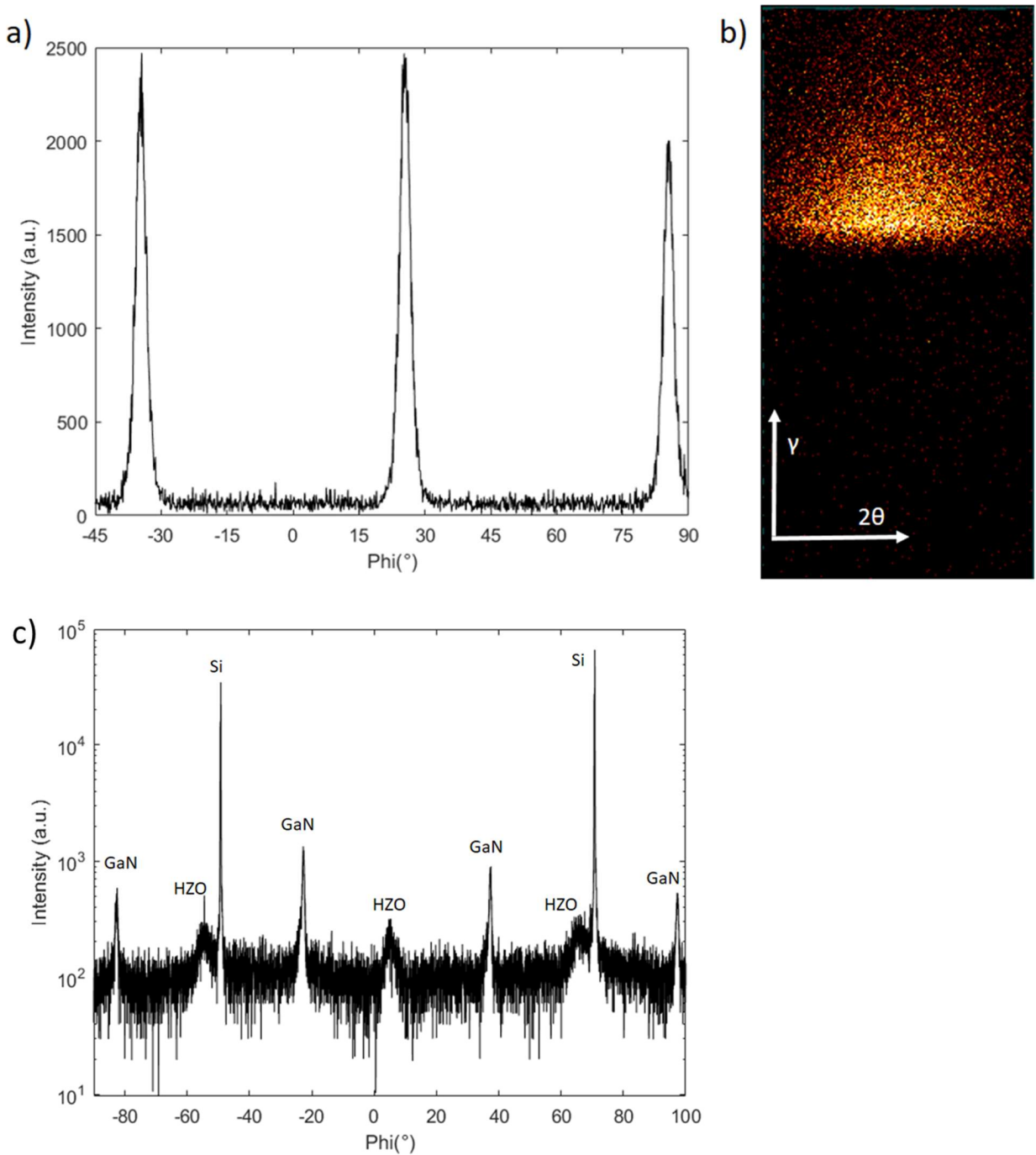

**Figure S3:** a) φ-scan of a 6 nm HZO film measured in IP-GIXRD geometry. The central pixel of the 2D detector was at 2θ = 50.8°, the incident angle was ω = 0.4°. b) A snapshot of the detector at ϕ = 85.5°. The directions 2θ and γ of the 2D detector are indicated in the figure. c) φ-scan of a 6 nm HZO film measured in IP-GIXRD geometry. The central pixel of the 2D detector was at 2θ =113.0°, the incident angle was ω = 0.4°.

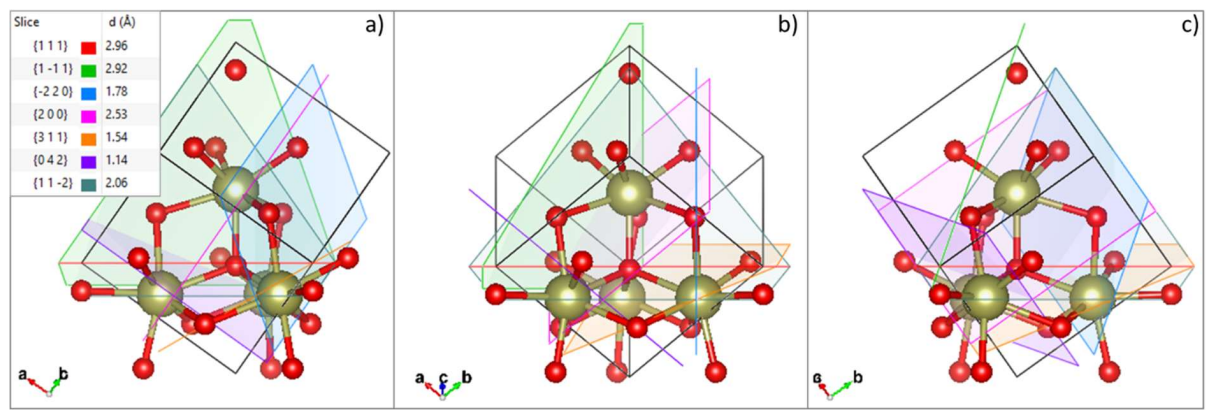

**Figure S4**: Model of the HZO unit cell with a R3 symmetry, a rhombohedral angle of α = 89.5° and a lattice parameter of a = 5.07 Å, projected along the (01-1) (a), (11-2) (b) and (10-1) (c) directions.